\documentclass[12pt]{article}
\usepackage{indentfirst}
\usepackage{graphicx}

\begin{document}

\author{R. R. R. Reis,  I. Waga,  M. O. Calv\~ao and S. E. Jor\'as\footnote{Present address:
State University of Santa Catarina (UDESC), Department of Physics,
P.\ O.\ Box 631,
CEP 89223-100, Joinville, SC , Brazil}  \\
{\small Federal University of Rio de Janeiro (UFRJ)}\\
{\small Institute of Physics, P. O Box 68528}\\
{\small CEP 21941-972 - Rio de Janeiro}\\
{\small Brazil}}
\title{Entropy perturbations in quartessence Chaplygin models}
\date{\today}
\maketitle

\begin{abstract}
We show that entropy perturbations can eliminate instabilities and
oscillations, in the mass power spectrum of the quartessence
Chaplygin models. Our results enlarge the current parameter space
of models compatible with large scale structure and cosmic
microwave background (CMB) observations.
\end{abstract}

\section{Introduction}

A spatially flat Universe whose evolution is driven by two
components, pressureless dark matter and negative pressure dark
energy, is currently considered the standard model of cosmology.
In spite of its success in explaining the majority of cosmological
observations, the exact nature of these components, which account
for more than $90\%$ of all the matter and energy content of the
Universe, is still a mystery. This situation spurred a search for
alternatives and, recently, considerable work has been devoted to
explore the possibility that dark matter and dark energy could be
distinct manifestations, at different scales, of a single
substance. This unifying dark matter-energy component is usually
referred to as UDM or quartessence \cite{makler03}. Two candidates
for such unification, that received a lot of attention lately, are
the quartessence Chaplygin fluid \cite{ kamenshchik01} and the
quartessence tachyonic field \cite{sen02}.

Of course, the first and most important task is to confront the
quartessence scenario with observations. In fact, it has been
shown that the quartessence Chaplygin model is not much
constrained by cosmological tests like SNeIa, lensing, age of the
Universe etc, that involve only the background metric
\cite{fabris02,makler03,avelino03,colistete03,dev03,silva03}. On
the other hand, tests that depend on the evolution of density
perturbations, narrow the parameter space so much, that only
generalized Chaplygin models whose behavior is close to the
$\Lambda$CDM limit are allowed
\cite{sandvik02,carturan02,bean03,amendola03}. However, in all the
analyses performed so far, it is assumed that pressure
perturbations are adiabatic. The main goal of this paper is to
show that if intrinsic entropy perturbations are allowed the
dynamics can change dramatically. A convenient choice for this
quantity will make a critical length \cite{sandvik02} in the
Chaplygin component vanish and, as a consequence, instabilities
and oscillations in this fluid are severely reduced. In particular
we show that models with negative values for the exponent $\alpha$
of the equation of state, which are discarded by the usual
analysis of adiabatic perturbations, may now be stable and, in
principle, compatible with CMB and large scale structure
observations.

\section{Evolution of Linear Perturbations}

The relativistic equations that govern the linear evolution of
scalar perturbations in a multi-component fluid, in the
synchronous gauge, are \cite{kodama, malik}
\begin{eqnarray} \displaystyle
\delta'_i + 3(c^{2}_{si}-w_i)\frac{a'}{a}\delta_i  & =  &
-(1+w_i)\left(kv_i+\frac{h'_L}{2}\right) -
3w_i\frac{a'}{a}\Gamma_i,
\label{eq1}  \\
v'_i + (1-3c^{2}_{si})\frac{a'}{a}v_i  &  =  &
\frac{c^{2}_{si}}{1+w_i}k\delta_i + \frac{w_i}{1+w_i}k\Gamma_i,
\label{eq2}  \\
h''_L + \frac{a'}{a}h'_L  &  =  &  -\sum_i(1+3c^{2}_{si})8\pi G
\rho_ia^2\delta_i - 24\pi Ga^2\sum_ip_i \Gamma_i,  \label{eq3}
\end{eqnarray}
where $\delta_i$, $v_i$, and $\Gamma_i$ are, respectively, the
density contrast, the velocity perturbation, and the entropy
perturbation of component $i$, $a$ is the scale factor, $h_L$ is
the trace of the metric perturbation, and $k$ is the comoving wave
number. For the sake of simplicity, we assume that both the
spatial curvature and the anisotropic pressure perturbation vanish
and that the energy-momentum tensor of each component is
separately conserved. In the equations above the prime denotes
derivative with respect to conformal time and, as usual,
$c_{si}^2=p'_i/\rho'_i$ and $w_i=p_i/\rho_i$ are, respectively,
the adiabatic sound speed and the equation of state parameter of
component $i$.

We will consider only two components in the universe: baryons
($p_b=0$) and a generalized Chaplygin fluid, whose homogeneous
equation of state is
\begin{equation}
p_{ch}=-M^{4(\alpha+1)}/\rho_{ch}^{\alpha}. \label{eq9}
\end{equation}
The separate conservation of the energy-momentum tensors leads to
\begin{equation}
\rho_b = \rho_{b0}a^{-3},  \label{eq10} \end{equation} and\footnote{%
Some authors (e.g. Makler et al. \cite{makler03}, and Sandvik et
al. \cite{sandvik02}) use a different notation for $A$. Their
``matter'' parameter is related to ours by the relation
$A=1-\Omega _{M}^{\ast }$.}
\begin{equation}
\rho_{ch} =
\rho_{ch0}\left[(1-A)a^{-3(\alpha+1)}+A\right]^{\frac{1}{\alpha+1}},
\qquad  A \equiv \frac{M^{4(\alpha+1)}}{\rho_{ch0}^{\alpha+1}},
\label{eq11}
\end{equation}
where the subscript ``0'' denotes quantities at the present time
and we choose the scale factor today equal to unity. It then turns
out that the equation of state parameter and the adiabatic sound
velocity, for the Chaplygin component, are given by
\begin{equation}
w_{ch}(a)= -\frac{Aa^{3(\alpha+1)}}{(1-A)+Aa^{3(\alpha+1)}},
\label{eq14}
\end{equation}
and
\begin{equation}
c^{2}_{sch} = - \alpha \;w_{ch}(a).  \label{eq12}  \\
\end{equation}

It is a good approximation to take $c^{2}_{sb}=0$ and we will also
assume that for baryons $\Gamma_b=0$. When the Chaplygin component
is considered, the instabilities and oscillations in the Chaplygin
mass power spectrum (and, to a lesser extent, also in the baryon
mass power spectrum), in the adiabatic case, have their origin in
a non-vanishing value of the right-hand side in Eq.\ (\ref{eq2})
\cite{sandvik02}. In this case, the effective sound speed and the
adiabatic sound speed are equal \cite{hu}. However, this may not
be true if entropy perturbations are present
\cite{reis03,balakin03,alam03}. Entropy perturbations are quite
common and, since the sound speeds of baryon and the Chaplygin
component are different, even for the so called ``adiabatic case''
there is, in fact, a relative entropy perturbation \cite{malik}.
We then impose that the right-hand side of Eq.\ (\ref{eq2})
vanishes, such that the effective sound speed for the Chaplygin
component is zero. This is equivalent to assuming, as an initial
condition, that $\delta p_{ch} =0$. It is straightforward to show
that $\delta p_{ch} =0$ implies $d\delta {p_{ch}}/dt =0$, such
that the condition is preserved \cite{abramo01}. With this choice
we can determine the intrinsic entropy perturbation $\Gamma_{ch}$
for the Chaplygin component, obtaining
\begin{equation}
\Gamma_{ch} =   -\frac{c^{2}_{sch}}{w_{ch}}\delta_{ch}=\alpha\;
\delta_{ch}.
\end{equation}

Since, for baryons $c_{sb}^2 =\Gamma_b=w_b=0$, it is easy to see
from Eq.\ (\ref{eq2}), that $v_b$ decays rapidly and, as a matter
of simplification, we take $v_b=0$. Also, from Eq.\ (\ref{eq1}),
we have, $\delta_b'=-h_L'/2$. Now, with
$\Omega_{b0}+\Omega_{ch0}=1$, Eqs.\ (\ref{eq1})-(\ref{eq3}) can be
recast as
\begin{equation}
\frac{d\delta_{ch}}{da} - 3w_{ch} (a)\frac{\delta_{ch}}{a} =
-(1+w_{ch} (a))\left(\frac{kv_{ch}}{\dot{a}a}
-\frac{d\delta_b}{da}\right),    \label{eq27}
\end{equation}
\begin{equation}
  \frac{dv_{ch}}{da} + (1+3\alpha w_{ch} (a))\frac{v_{ch}}{a} = 0,
   \label{eq28}
\end{equation}
and
\begin{equation}
  \frac{d^2\delta_b}{da^2} + \left(\frac{2}{a} +
\frac{\ddot{a}}{\dot{a}^2}\right)\frac{d\delta_b}{da}  =
\frac{3H_{0}^{2}}{2\dot{a}^2}\left\{\frac{\Omega_{b0}\delta_b}{a^3}
+(1-\Omega_{b0})\delta_{ch}\left[\frac{(1-A)}{a^{3(\alpha+1)}}
+A\right]^{\frac{1}{\alpha+1}}\right\},    \label{eq29}
\end{equation}
where $\dot{a}$ and $\ddot{a}$ are given by
\begin{eqnarray}
\dot{a}=H_0\left\{\frac{\Omega_{b0}}{a}
+(1-\Omega_{b0})\left[Aa^{2(\alpha+1)}
+\frac{(1-A)}{a^{\alpha+1}}\right]^{\frac{1}{\alpha+1}}\right\}^{\frac{1}{2}},
\label{eq25}  \\
\ddot{a}=-\frac{H_{0}^{2}}{2}\left\{\frac{\Omega_{b0}}{a^2}
+(1-\Omega_{b0})(1+3w_{ch} (a))\left[Aa^{\alpha+1}
+\frac{(1-A)}{a^{2(\alpha+1)}}\right]^{\frac{1}{\alpha+1}}\right\}.
\label{eq26}
\end{eqnarray}

\section{Discussion}

In our numerical calculations we evolved Eqs.\ (\ref{eq27}),
(\ref{eq28}) and (\ref{eq29}) from $z=500$ to $z=0$ to obtain the
power spectra. At $z=500$ the Chaplygin fluid behaves like CDM
and, to take this into account, proper initial conditions were
established; we assumed $v_{ch}=0$, a scale invariant primordial
spectrum and used the BBKS transfer function, \cite{bardeen86}
with the following effective shape parameter
\cite{sugiyama95},\cite{beca03},
\begin{equation}
\Gamma_{eff}= (\Omega_{b0}+(1-\Omega_{b0})(1-A)^{1/(1+\alpha)})h\;
\exp\left(-\Omega_{b0}-\frac{\sqrt{2h}\;\Omega_{b0}}
{\Omega_{b0}+(1-\Omega_{b0})(1-A)^{1/(1+\alpha)}}\right).
\end{equation}

In Fig.\ 1 we present the Chaplygin mass power spectra, in the
adiabatic case ($\Gamma_{ch}=0$), for $\alpha = - 10^{-4},
-10^{-5}, 0, 10^{-5}$ and $10^{-4}$ (from top to bottom), $h=0.7$,
$\Omega_{b0}=0.04$ and we choose $A$ such that all models have
$\Gamma_{eff}=0.18$. The normalization of the mass power spectra
is arbitrarily fixed at $k=0.01$ h Mpc$^{-1}$. For larger absolute
values of $\alpha$ the oscillations and instabilities are much
stronger. The data points are the power spectrum of the 2dF galaxy
redshift survey as compiled in \cite{tegmark02}. In Fig.\ 2, for
the same parameters, we show the baryon power spectra. It is clear
that, for baryons, the oscillations and, to a lesser extent,
instabilities (that occur for $\alpha < 0$) are quite reduced
\cite{beca03}.

In Fig.\ 3 we show the mass power spectra for the Chaplygin fluid
in the case $\Gamma_{ch}= \alpha \;\delta_{ch}$ ($\delta
p_{ch}=0$). We considered $\alpha = - 0.6, -0.3, 0, 0.3, 0.6$ and
$1$, $h=0.7$, $\Omega_{b0}=0.04$ and, again, $A$ such that all
models have $\Gamma_{eff}=0.18$. The degeneracy is clear (only one
curve is visually discernible) and, as remarked before, at linear
scales, the oscillations and instabilities are absent. Note that
models with $\alpha < 0$ may now be in agreement with
observations. The baryon mass power spectra curves essentially
coincide with these ones. In Fig.\ 4 we show the mass power
spectra for the Chaplygin component, in the non-adiabatic case,
for the same values of $\alpha$, $h$ and $\Omega_{b0}$ as above,
but now we fixed the parameter $A=0.72$, such that $\Gamma_{eff}$
is different for each model. The degeneracy has now been broken.
Again the baryon mass power spectra curves are essentially
identical.

In the standard cosmological model ($\Lambda$CDM and QCDM), two
mysterious components, dark matter and dark energy, are predicated
to explain two different phenomena: clustering of matter and
cosmic acceleration. In spite of the standard cosmological model's
success, so far it has not been proven that dark matter and dark
energy are, in fact, two distinct substances. A first attempt has
been presented in \cite{sandvik02}. In \cite{amendola03} it has
been shown that CMB observations strongly constrain the parameter
space of quartessence Chaplyging models, leaving little room for
models dissimilar to $\Lambda$CDM. All these results are based on
the assumption that there is no intrinsic entropy perturbation in
the Chaplygin component. In this work we argue that if
$\Gamma_{ch}=\alpha \delta_{ch}$, or, equivalently, $\delta
p_{ch}=0$ as an initial condition, oscillations and instabilities,
present in the adiabatic case, will disappear and, consequently,
the parameter space will be enlarged (for instance, now $\alpha<0$
should be included in the analyses). We believe that the game is
not over or in its final round for these models, perhaps it has
just started. Anyway, our results suggest that they deserve
further investigation.

\bigskip {\bf Acknowledgments}

We thank helpful conversations with Raul Abramo, Fabio Finelli,
Martin Makler and Max Tegmark. The authors would like to thank the
Brazilian research agencies CAPES, CNPq, FAPERJ and Funda\c{c}\~ao
Jos\'e Bonif\'acio for financial support.

\begin{figure}\centering \hspace*{-0.8in}
\includegraphics[height= 12 cm,width=12cm]{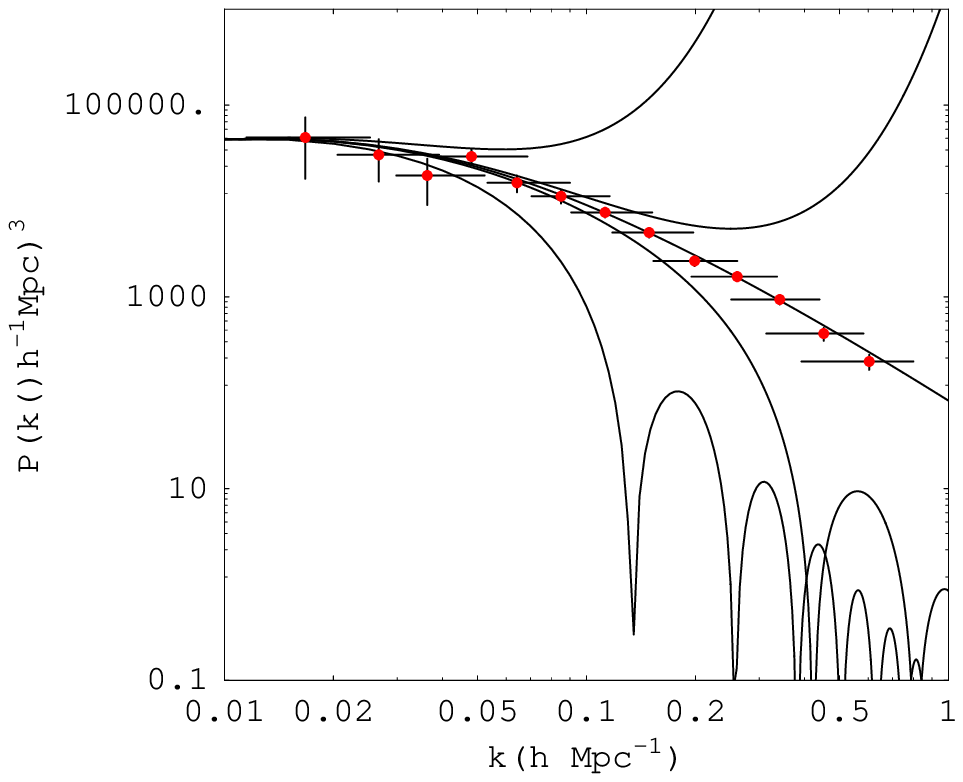}
\caption{Chaplygin mass power spectra, in the adiabatic case, for
$\alpha = - 10^{-4}, -10^{-5}, 0, 10^{-5}$ and $10^{-4}$ (from top
to bottom), $h=0.7$, $\Omega_{b0}=0.04$ and $A$ chosen such that
all models have $\Gamma_{eff}=0.18$. The data points are the power
spectrum of the 2dF galaxy redshift survey as compiled in
\cite{tegmark02}. }
\end{figure}

\begin{figure}\centering \hspace*{-0.8in}
\includegraphics[height= 12 cm,width=12cm]{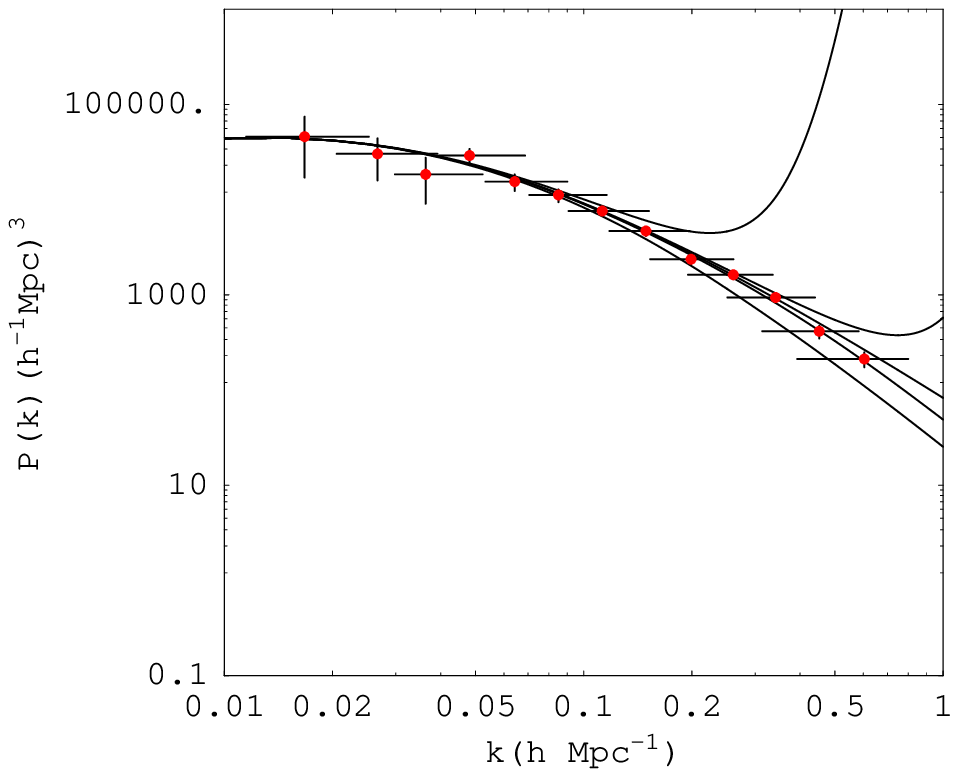}
\caption{Baryon mass power spectra, in the adiabatic case, for
$\alpha = - 10^{-4}, -10^{-5}, 0, 10^{-5}$ and $10^{-4}$ (from top
to bottom), $h=0.7$, $\Omega_{b0}=0.04$ and $A$ chosen such that
all models have $\Gamma_{eff}=0.18$. The data points are the power
spectrum of the 2dF galaxy redshift survey as compiled in
\cite{tegmark02}. }
\end{figure}

\begin{figure}\centering \hspace*{-0.8in}
\includegraphics[height= 12 cm,width=12cm]{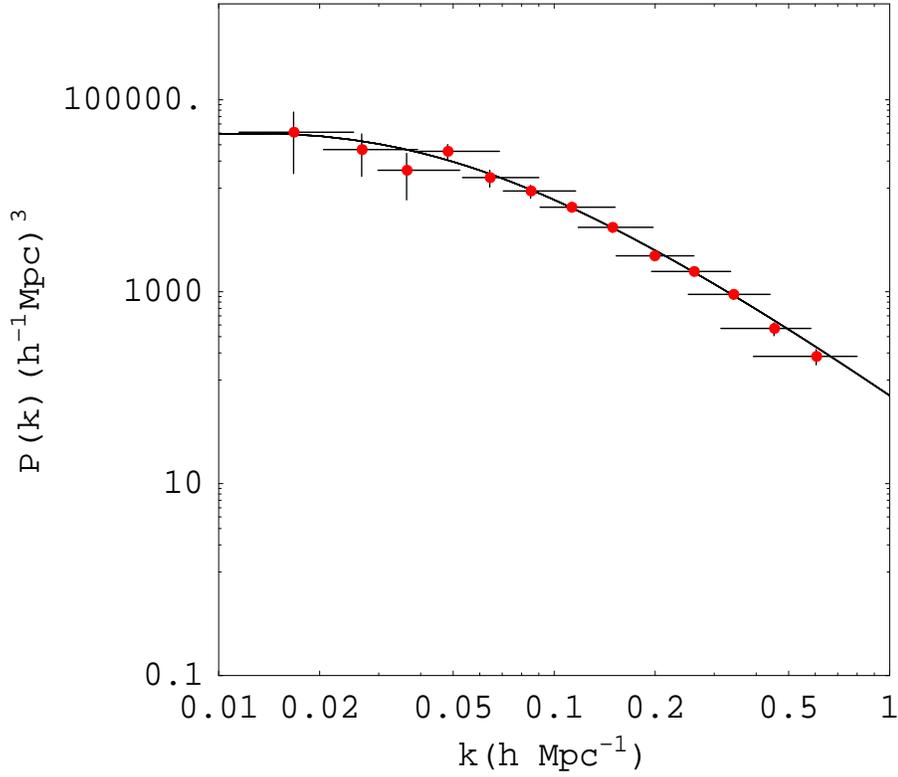}
\caption{Chaplygin (and baryon) mass power spectra, in the
non-adiabatic case, for $\alpha = - 0.6, -0.3, 0, 0.3, 0.6$ and
$1$, $h=0.7$, $\Omega_{b0}=0.04$ and $A$ chosen such that all
models have $\Gamma_{eff}=0.18$. The data points are the power
spectrum of the 2dF galaxy redshift survey as compiled in
\cite{tegmark02}. }
\end{figure}
\begin{figure}\centering \hspace*{-0.8in}
\includegraphics[height= 12 cm,width=12cm]{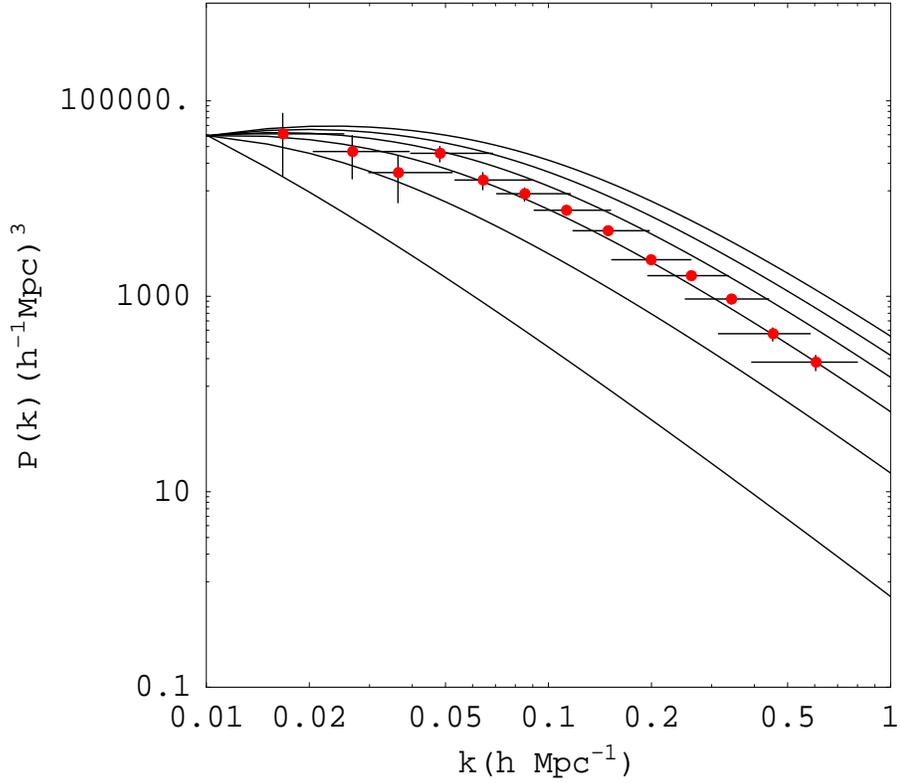}
\caption{Chaplygin (and baryon) mass power spectra, in the
non-adiabatic case, for $\alpha = - 0.6, -0.3, 0, 0.3, 0.6$ and
$1$, (from bottom to top), $h=0.7$, $\Omega_{b0}=0.04$ and
$A=0.72$. The data points are the power spectrum of the 2dF galaxy
redshift survey as compiled in \cite{tegmark02}. }
\end{figure}
\end{document}